\documentclass{article}

\usepackage{graphicx}

\usepackage{amsfonts}
\usepackage{natbib}
\usepackage{epsfig}
\usepackage{amsfonts} 
\usepackage{amsmath}
\usepackage{amssymb}
\usepackage{multirow}

\newcommand{\vecx}{\mathbf{x}}

\newcommand{\wg}{\mathcal{W}}

\begin{document}

\setcounter{page}{1}

\title{Variable Selection for Clustering and Classification\footnote{A version of this manuscript will appear in the \emph{Journal of Classification}; once available, a DOI will be provided.}}

\author{Jeffrey L. Andrews\footnote{Corresponding author. E-mail: {andrewsj@uoguelph.ca}. Telephone: +1-519-824-4120 ext.\ 56558} \ and Paul D. McNicholas \\{\small Department of Mathematics \& Statistics, University of Guelph, Guelph, ON, Canada,~N1G2W1.}}

\date{}

\maketitle

\begin{abstract}
As data sets continue to grow in size and complexity, effective and efficient techniques are needed to target important features in the variable space. Many of the variable selection techniques that are commonly used alongside clustering algorithms are based upon determining the best variable subspace according to model fitting in a stepwise manner. These techniques are often computationally intensive and can require extended periods of time to run; in fact, some are prohibitively computationally expensive for high-dimensional data. In this paper, a novel variable selection technique is introduced for use in clustering and classification analyses that is both intuitive and computationally efficient. We focus largely on applications in mixture model-based learning, but the technique could be adapted for use with various other clustering/classification methods. Our approach is illustrated on both simulated and real data, highlighted by contrasting its performance with that of other comparable variable selection techniques on the real data sets.
\\
\vspace{0.2in}

\end{abstract}

\section{Introduction}
\label{sec:introduction}
Variable selection is an important feature of many types of statistical analyses, including clustering and classification. The use of variable selection techniques can facilitate both model fitting and the interpretation of results. With the continued growth in data dimensionality, the importance of efficient variable selection techniques is increasing. In this paper, we put forward a flexible variable selection technique that can be used in unsupervised, semi-supervised, or fully supervised classification contexts. We focus largely on the unsupervised format (i.e., clustering) and specifically the usage of model-based techniques to guide the algorithm. However, the algorithm's applicability under a classification format will be briefly illustrated in Section~\ref{sec:sim}.

Model-based clustering is often used on high-dimensional data sets, such as those found in the field of bioinformatics. Though clustering on very large data sets is possible, it is difficult to execute for a number of reasons. The most obvious reason is time: as the dimensionality of the data increases, the number of parameters requiring estimation increases, often in a quadratic fashion. Of more importance, however, is that the human brain (accustomed to three-dimensions visually, plus a few other senses) is not prepared to understand dimensionality that can run well into the thousands. Thus, to facilitate interpretation of high-dimensional data sets, determining which variables are most active in cluster formation is important. A final consideration is the cost in creating high-dimensional data sets, which can be enormous; knowing which variables are important for differentiating between groups-of-interest can save both time and money.

Although algorithm efficiency is certainly a worthwhile reason in its own right, variable selection techniques can drastically improve clustering performance. This is achieved through eliminating noisy variables that can cloud the clustering algorithm's ability to distinguish groups. Unfortunately, variable selection techniques do not necessarily improve clustering performance; it is, therefore, important that an inferior reduced-variable solution is not chosen over a solution on the full variable set (cf.\ Section~\ref{sec:app}).

In Section~\ref{sec:back}, we conduct a short review of comparable variable selection techniques and other relevant background material. Then we discuss our methodology (Section~\ref{sec:meth}), before running simulations (Section~\ref{sec:sim}) and real-data examples (Section~\ref{sec:app}). Finally,  we conclude with a summary and suggestions for future work (Section~\ref{sec:disc}). 

\section{Background}\label{sec:back}
A number of dimensionality reduction techniques are available to researchers interested in clustering data sets. For the purposes of microarray data sets, the {\tt select-genes} procedure from the EMMIX-GENE software \citep{mclachlan02} fits multi-component mixture models to each variable and then calculates the likelihood ratio test statistic between these and the one-component model. Unfortunately, fitting mixture models to each individual variable is time consuming and, by relying on random initializations, {\tt select-genes} can be inconsistent.

Another variable selection technique is given by \cite{raftery06}, whereby multiple models from the MCLUST family are compared using approximate Bayes factors \citep{kass95}. This variable selection technique is readily available via the {\tt clustvarsel} package \citep{dean06b} in {\sf R} \citep{R12}. However, because the number of parameters that require estimation in some of the MCLUST models is quadratic in data-dimensionality, the {\tt clustvarsel} package can be very slow in high-dimensions. Furthermore, the application of {\tt clustvarsel} can sometimes lead to inferior results when compared to the use of {\tt mclust} alone \citep[cf.][]{mcnicholas08}. A related approach, denoted {\tt selvarclust}, is taken by \cite{maugis09}, where the assumptions on the role of variables are relaxed with the potential benefit of avoiding the over-penalization of independent variables.

In addition to these procedures, a number of implicit and explicit variable selection procedures are built into model-based clustering algorithms. Implicit variable selection procedures include approaches such as mixtures of factor analyzers \citep[cf.][]{ghahramani97,tipping99b,mclachlan00a, mcnicholas08,mcnicholas10d,andrews11c,andrews11b,montanari10,viroli10}. An explicit dimensionality reduction approach is taken in some recent work by \cite{scrucca09} and \cite{bouveyron11}; the latter also gives a summary of  other work in the area of dimensionality reduction with respect to clustering. Outside of model-based methods, \cite{witten10} recently introduced a dimensionality reduction technique that is based on $k$-means; we provide a comparison to this technique in Section~\ref{sec:skm}.

For the purposes of variable selection within a clustering context, the desire is to find variables that show differentiation between the \emph{a priori} unknown groups and eliminate variables that do not. The variable selection method introduced herein seeks precisely this, and is flexible enough to implement using a variety of clustering/classification techniques.

\section{Methodology}\label{sec:meth}
\subsection{Introduction}
Variable selection for clustering and classification (VSCC) is intended to find the variables that simultaneously minimize the `within-group' variance and maximize the `between-group' variance. The combination of these two criteria will give variables that best show separation between the desired groups. Note that the within-group variance for each variable $j=1,\ldots,p$ can be written as
\begin{equation*}
\wg_j = \frac{\sum_{g=1}^G \sum_{i=1}^{n} z_{ig} (x_{ij} - \mu_{gj})^2}{n},
\end{equation*}
where $x_{ij}$ is observation $i$ on variable $j$, $\mu_{gj}$ is the mean of variable~$j$ in group~$g$, $n$ is the number of observations, and $z_{ig}$ is a group membership indicator variable defined so that
\begin{equation*}
z_{ig} = \left\{ 
\begin{array}{l}
1 \quad \mbox{if observation $\vecx_i=(x_{i1},\ldots,x_{ip})$ belongs to cluster $g$,}  \\
0 \quad \mbox{otherwise.}
\end{array}
\right.
\end{equation*}
The leftover variance within variable $j$ not accounted for by $\wg_j$, or $\sigma^2_j - \wg_j$ in the common notation, is then a measure of the variance between groups. In general, calculation of this value will be necessary. However, if the data have been standardized to have equal variance across variables, then any variable minimizing the within-group variance is also maximizing the leftover variance. 

The VSCC method utilized in this article will be applied to data that have been standardized to have mean~0 and variance~1 and, as such, calculation of the $\wg_j$ is sufficient. In addition to the variance calculations, our method uses the correlation between variables; we let $\rho_{ij}$ denote the correlation between variables~$i$ and~$j$. The actual implementation of VSCC procedures depends on the form of the data; in our analyses, we consider examples where no memberships are known (clustering) as well as where some observations have known membership (classification).  Specifics regarding the implementation of the VSCC algorithm under each format can be found in Sections~\ref{sec:clus} and \ref{sec:clas}. In the sections that immediately follow, we motivate and describe the VSCC algorithm.

\subsection{A Motivating Example}
VSCC will proceed in a step-wise fashion after calculating the within-group variances. The first variable selected is the variable with the minimum $\wg_j$. One way to select from the remaining variables is by using simple, user-specified thresholding. For example, by sorting the $\wg_j$ in ascending order we could consider each variable in a step-wise manner and select those variables with $\wg_j$ less than some value $w$ where all  $|\rho_{jr}|$ are also less than some value $c$, $\forall$ $r \in V$; here, $V$ is the set of previously selected variables. While this approach could be useful, it requires the user to adjust the algorithm to maximize its effectiveness. 

An additional concern behind this type of selection criterion can be shown via a simple example. Consider a three-dimensional data set where the first variable minimizes $\wg_j$ and so is already selected. Suppose that the remaining two variables can be summarized as follows.
\begin{itemize}
\item Variable 2: $\wg_2=0.6$ and $|\rho_{12}|=0.75$.
\item Variable 3: $\wg_3=0.2$ and $|\rho_{13}|=0.75$.
\end{itemize} 
Suppose we simply use the thresholds described previously. In the current example, if the correlation threshold was set at $c=0.70$ then both variables would be considered equally `bad' and neither would be selected. However, if the correlation threshold was set at $c=0.80$, and assuming both that $|\rho_{23}|<0.8$ and $w> 0.6$, then both variables would be selected. 
Under the argument that within-group variance is our primary concern and correlation is a secondary concern, we propose that, in this scenario, retaining Variable~3 and eliminating Variable~2 would be desirable. To this end, we need to go beyond simple, user-specified thresholding.

\subsection{The VSCC Method}
We have illustrated a desire for a sliding correlation threshold that is more forgiving for small values of $\wg_j$ and more stringent for larger values. Thus, we seek to define a relationship between the within-group variance and between-variable correlation that properly expresses this goal. As a first attempt, we consider a linear relationship between the two quantities. Let $V$ represent the space of currently selected variables, then we select variable $j$ if for all $r \in V$, $$|\rho_{jr}| < 1 - \wg_j.$$ Other potential relationships will be discussed shortly, but utilizing this relationship we can write the VSCC algorithm as follows:
\begin{enumerate}
\item Calculate within-group variances $\wg_j$.
\item Sort $\wg_j$ in ascending order, denote this sorted list ${\bf W}_s$. 
\item $\mathcal{W}_1$ minimizes ${\bf W}_s$ and is automatically selected and placed into the set of selected variables $V$. Set count $k=2$.
\item If $|\rho_{kr}| < 1 - W_k$, for all $r \in V$, variable $s=k$ is placed into $V$.
\item If $k<p$, set $k=k+1$ and return to Step 4. Else end algorithm.
\end{enumerate}

\begin{figure}[h]
\begin{center}
\includegraphics[width=5in]{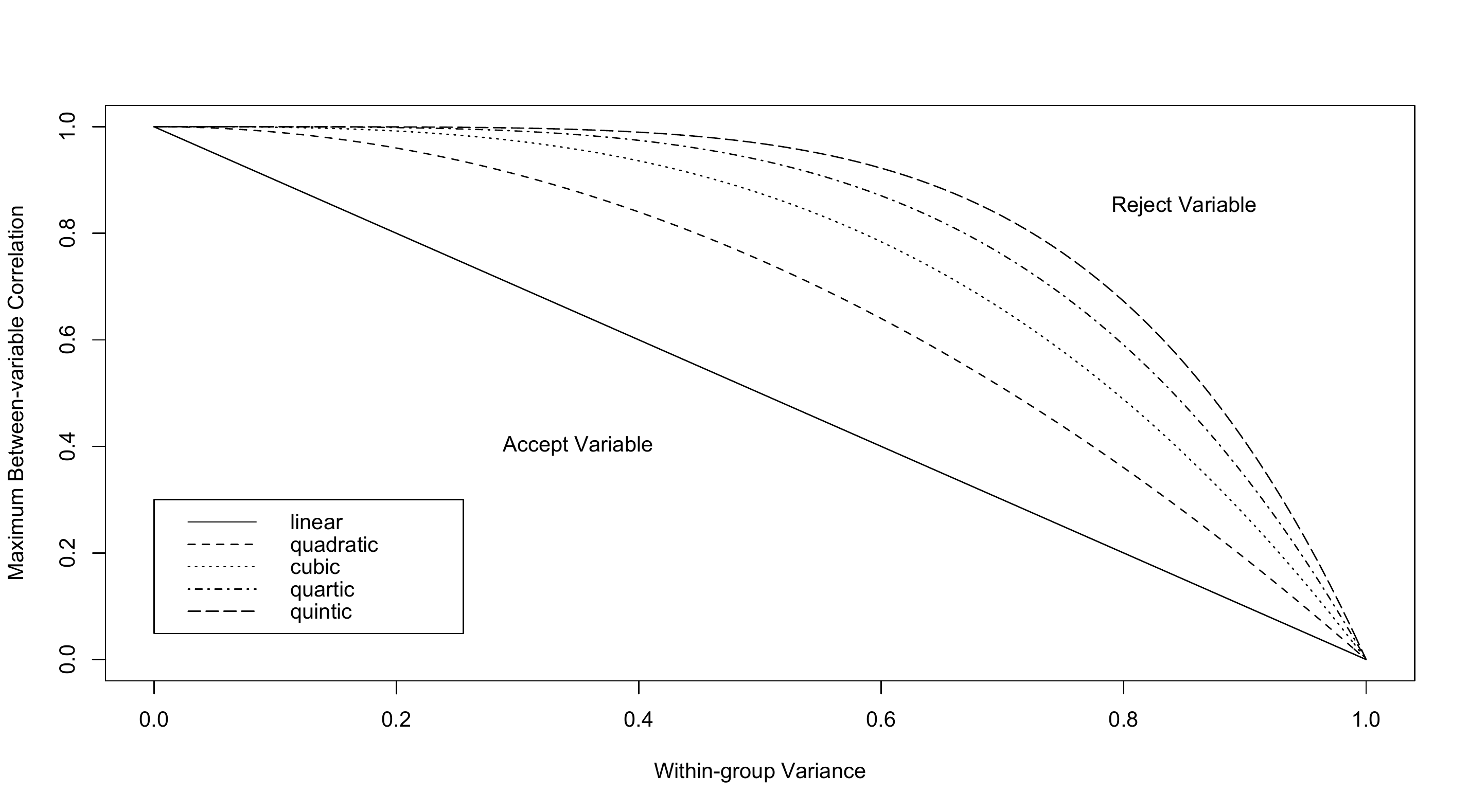}
\caption{Graphical representation of the selection criteria considered for implementation into Step 4 of the VSCC algorithm.}
\label{fig:rel}
\end{center}
\end{figure}

The linear relationship defined in Step 4 of the VSCC algorithm might be too strong a criterion. For instance, a variable with within-group variation of 0.25 and correlation of 0.76 with one of the previously selected variables would be rejected. Given the interval that correlation values (and the $\wg_j$ when the data are standardized) will lie on, a simple fix is to consider relationships of order greater than one (Table~\ref{ta:rel}); a visualization of these criteria is given in Figure~\ref{fig:rel}. 
\begin{table}[htb]
\caption{List of variance-correlation relationships considered for implementation into Step 4 of the VSCC algorithm.}
\label{ta:rel}
\begin{center}
\begin{tabular}{ll}
\hline
Linear & $|\rho_{kr}| < 1 - W_k$ \\[+2pt]
Quadratic & $|\rho_{kr}| < 1 - W_k^2$ \\[+2pt]
Cubic & $|\rho_{kr}| < 1 - W_k^3$ \\[+2pt]
Quartic & $|\rho_{kr}| < 1 - W_k^4$ \\[+2pt]
Quintic & $|\rho_{kr}| < 1 - W_k^5$ \\
\hline
\end{tabular}
\end{center}
\end{table} 

Many different relationships could be defined between the within-group variance and the between-variable correlation. However, many of these relationships would be intuitively silly. For example, any relationship that will allow a variable with $\wg_j=1$ to be selected should not be allowed. Also, any relationship that results in impossible values (those outside of the interval [0,1]) need not be considered. Obviously, piecewise relationships could be considered that solve some of these issues for more complicated relationships. However, we consider that the relationships in Table~\ref{ta:rel} constitute a relatively thorough, common-sense handling of the issue of variance-correlation relationships.

\subsection{Subset Selection}\label{sec:rel}
Using multiple criteria for selecting variables will naturally lead to multiple subsets of variables --- as many as five solutions under the current relationship structure. Under a clustering framework, one must define a method for choosing between these subsets without specific knowledge of which subset produces the best classifier. 
One could develop a Bayes factors framework similar to that used by {\tt clustvarsel} \citep{raftery06} to compare variable subsets. However, this approach is complicated by the fact that subsets will not necessarily differ by only one variable, as would happen in a truly step-wise approach. Instead, we introduce a novel approach to selecting variable subsets that relies on one of the major strengths of model-based clustering/classification: measuring the uncertainty of the classification.

The uncertainty for each observation is found simply through the fuzzy classification matrix; i.e., the $n\times G$ matrix containing the $\hat z_{ig}$. Each $\hat z_{ig}$ element of this matrix is a measure of the strength of evidence indicating observation~$i$ belongs to group~$g$. For well-defined clusters, the $\hat z_{ig}$ will all be approximately equal to 0, with one entry per row $i$ being approximately equal to 1. We take the uncertainty to be the sum of all the $\hat z_{ig}$ entries, except the $\max_g\{ \hat z_{ig}\}$ entries. This can be expressed as
$\sum_{i=1}^n \sum_{g=1}^G \hat z_{ig} -  \sum_{i=1}^n \max_g \{\hat z_{ig}\}$
or equivalently as $n - \sum_{i=1}^n \max_g \{\hat z_{ig}\}$.

Selecting the variable subset that minimizes the uncertainty in the classification suggests we will be selecting the variables that produce the strongest group structure, so there is some intuitive appeal. In some ways, this is a large departure from information-based criteria; however, the uncertainty is used in the calculation of the integrated completed likelihood by \cite{biernacki00}, which marries the uncertainty and the Bayesian information criterion \citep[BIC;][]{schwarz78}. The efficacy of using the uncertainty as a relationship selection criterion will be shown in Sections~\ref{sec:sim} and \ref{sec:app}.

The authors note that one concern about using the uncertainty calculations is that, by definition, the uncertainty for any $G=1$ solution will be 0. It is considered a strength within the clustering field that model-based clustering using the BIC can consider $G=1$ as a solution and inform the user that there are, in fact, no groups in the data. Unfortunately, given a $G=1$ solution, the VSCC algorithm cannot be computed as described in this paper. As such, it is an implicit assumption by even running VSCC that $G>1$ groups exist. Tying into this assumption, it is therefore somewhat reasonable to ignore any variable subsets that produce $G=1$ as a solution. We recognize this is an unfortunate consequence of utilizing model uncertainty as a subset selection device, and we leave this matter as the subject of future research.

Philosophically, it makes sense to approach variable selection under a ``do no harm'' mentality. In this vein, note that because we can calculate the uncertainty from the original (non-feature reduced) data set, this solution can be considered as part of the variable selection process. In other words, under VSCC we can select the full data set rather than a reduced set if the full data set results in the minimum uncertainty; we note that for illustrative reasons we will ignore this ability during the simulations in Section~\ref{sec:sim}.

\subsection{Clustering}\label{sec:clus}

In a clustering scenario, the values of indicator variables $z_{ig}$ are unknown for all $i=1,\ldots,n$. The VSCC method needs these variables to compute the $\wg_j$. However, an initial run of clustering can be used to give `good' estimates of the component memberships $\hat z_{ig}$; any clustering approach could be used (e.g., agglomerative hierarchical, model-based, $k$-means). 

Herein, we focus on the use of a model-based clustering procedure to initialize VSCC, specifically the {\tt mclust} algorithm \citep{fraley02a, fraley03, fraley06}. Note that the applications in Section~\ref{sec:app} will incorporate `hard' (0s and 1s in the context of $\hat z_{ig}$) initializations from a model-based clustering technique, though `soft' (`fuzzy' or probabilistic) classifications could be easily incorporated into the procedure instead. Under the clustering format, the algorithm runs as follows:
\begin{enumerate}
\item Perform {\tt mclust} under default settings.
\item Use the resultant (hard) $\hat z_{ig}$ to initialize VSCC.
\item Perform {\tt mclust} on the (up to) five variable subsets given by VSCC, selecting the best model via the BIC in each case.
\item Select the best variable subset according to the total model uncertainty, and report the results from {\tt mclust} on that subset.
\end{enumerate}

\subsection{Classification}\label{sec:clas}
In a classification scenario, a subset of the $z_{ig}$ is known and can be utilized by the VSCC method to compute the $\wg_j$ in two potential ways. The first option is to use only the known $z_{ig}$ to calculate the $\wg_j$. The other option is to calculate the $\wg_j$ in a more semi-supervised format, where an original classification algorithm is run to find good estimates of the unknown $\hat z_{ig}$. The algorithms for both options follow.

\subsubsection{Supervised Algorithm}\label{sec:sup}
\begin{enumerate}
\item Use only the known $z_{ig}$ to initialize VSCC.
\item Perform model-based classification on the (up to) five variable subsets given by VSCC, selecting the best model via the BIC in each case.
\item Select the best variable subset according to the total model uncertainty, and report the results from model-based classification on that subset.
\end{enumerate}

\subsubsection{Semi-Supervised Algorithm} 
\begin{enumerate}
\item Perform model-based classification to estimate the unknown $\hat z_{ig}$.
\item Use both the known $z_{ig}$ and the (hard) estimated $\hat z_{ig}$ to initialize VSCC.
\item Perform model-based classification on the (up to) five variable subsets given by VSCC, selecting the best model via the BIC in each case.
\item Select the best variable subset according to the total model uncertainty, and report the results from model-based classification on that subset.
\end{enumerate}

\subsection{Clustering/Classification Performance}
The performance of a clustering algorithm, with respect to known groups present in the data, can be measured in a number of ways. Misclassification rates are often used, but this measure cannot be meaningfully interpreted unless we know the correct number of groups or the clustering algorithm chooses the correct number of groups, which is not always the case. An alternative is to use the Rand index \citep{rand71}, which is calculated as the number of pairwise agreements (between the estimated and known groups) divided by the number of pairs. Because the Rand index is tricky to interpret especially for lower values, the adjusted Rand index (ARI) was introduced by \cite{hubert85}. It essentially accounts for the fact that two random groupings will have some pairwise agreements, thus making the ARI equal to 0 for random clustering and 1 for perfect clustering.

\section{Simulations}\label{sec:sim}
To determine the performance of VSCC under a variety of scenarios, we introduce several simulation studies. The {\tt clusterGeneration} package \citep{qiu06} from {\sf R} is used to simulate data sets.

\subsection{Increased Dimension}
Herein we investigate how VSCC reacts to increased dimension. The {\tt genRandomClust} function from {\tt clusterGeneration} is used to generate data sets with four groups, with between 100 to 150 observations per group and where {\tt sepVal}$=0.7$ (well separated groups). We generate data sets of dimension 45, 90, 120, and 150, each containing 33\% noisy variables. A total of 250 replicates of each data set are generated and analyzed using VSCC and {\tt mclust} (under default settings). Summary results are given in Table~\ref{ta:incdall}.

\begin{table}[h]
\begin{center}
\caption{Summary of results from {\tt mclust} and VSCC on the increased dimension simulations (250 runs per dimension size).}
\label{ta:incdall}
\begin{tabular}{lrrrr}
\hline
&$d=45$&$d=90$&$d=120$&$d=150$\\
\hline
{\tt mclust} Mean ARI&0.79&0.36&0.30&0.23\\
{\tt mclust} SD ARI&0.15&0.14&0.17&0.17\\
{\tt mclust} Avg Runtime (sec)&5.16&14.74&72.01&157.82\\
\\
VSCC Mean ARI&0.99&0.84&0.76&0.57\\
VSCC SD ARI&0.03&0.20&0.30&0.33\\
VSCC Avg Runtime (sec)&25.95&42.89&160.04&265.62\\

\hline
\end{tabular}
\end{center}
\end{table}

VSCC performs better than {\tt mclust} alone on all dimension sizes considered, according to mean ARI. We do, however, note an increase in the standard deviation of the ARI as the dimension size increases. This is due, in large part, to an increased number of $G=1$ solutions given by {\tt mclust} for an initialization (which is counted as an ARI of 0 for both {\tt mclust} and VSCC). Because {\tt mclust} performance is on average closer to 0, these $G=1$ examples affect its standard deviation to a lesser extent.

Note also the increase in runtime for both procedures. Keep in mind that VSCC runs {\tt mclust} once on the full data set, and then up to five times on reduced-variable data sets. Thus, a large savings in computation time could be achieved by at least initializing VSCC using a faster clustering technique (e.g., $k$-means). To illustrate this point, if the initializations were given to VSCC `free-of-charge' for the $d=150$ simulations, VSCC's average runtime would be merely 98.8 seconds.

\subsection{Increased Number of Groups}
In this simulation, we investigate how VSCC reacts to different numbers of groups present in the data. Once again, the {\tt genRandomClust} function is used to generate 250 replicates of each data set. In this study, we simulate under mostly default conditions --- which includes {\tt sepVal}=0.01, or not well separated groups --- with 10 noisy and 10 non-noisy variables in each data set. Importantly, we generate data sets for each of $G=2,4,6,8,15,20$ that are then analyzed using VSCC and {\tt mclust} (under default settings except for $G=15$ and $G=20$ data sets, where {\tt mclust} is forced to consider $G=10,\ldots,20$). Summary results are given in Table~\ref{ta:incgall}. We also provide Table~\ref{ta:g2} for more in-depth details on the $G=2$ simulation to illustrate the specific variance-correlation relationships as well as the performance of choosing relationships via the total model uncertainty.
\begin{table}[h]
\begin{center}
\caption{Summary of results from {\tt mclust} and VSCC on the varied number of groups simulations (250 runs per group structure).}
\label{ta:incgall}
\begin{tabular}{lrrrrrrr}

\hline
&$G=2$&$G=4$&$G=6$&$G=8$&$G=15$&$G=20$\\
\hline
{\tt mclust} Mean ARI&0.88&0.79&0.80&0.75&0.74&0.73\\
{\tt mclust} SD ARI&0.08&0.17&0.11&0.13&0.06&0.05\\
{\tt mclust} Avg.\ Runtime (sec.)&2.47&6.89&10.76&16.59&97.51&197.31\\
\\
VSCC Mean ARI&0.89&0.82&0.85&0.83&0.84&0.82\\
VSCC SD ARI&0.06&0.17&0.08&0.08&0.03&0.03\\
VSCC Avg.\ Runtime (sec.)&8.95&18.21&22.87&31.82&171.39&302.77\\

\hline
\end{tabular}
\end{center}
\end{table}
\begin{table}[b]
\begin{center}
\caption{Detailed results from {\tt mclust} and VSCC on the 250 $G=2$ simulations (10 noisy variables and 10 non-noisy variables). `Mode \# Vars' includes the number of occurrences in parentheses.}
\label{ta:g2}
\begin{tabular}{lrrrrrr}
\hline
Analysis&Mean ARI&SD ARI&Mode \#Vars&Mean Unc.& SD Unc.\\
\hline
{\tt mclust}&0.88&0.08&20 (250)&5.88&3.81\\
VSCC Linear&0.40&0.30&2 (194)&14.45&13.27\\
VSCC Quadratic&0.81&0.09&3 (75)&9.67&5.26\\
VSCC Cubic&0.86&0.08&6 (74)&7.32&4.45\\
VSCC Quartic&0.87&0.09&6 (82)&6.67&4.29\\
VSCC Quintic&0.88&0.09&6 (71)&6.27&4.21\\
VSCC (min unc.)&0.89&0.06&6 (75)&5.73&2.86\\
\hline
\end{tabular}
\end{center}
\end{table}

As the number of groups increases, the general trend for {\tt mclust} is a reduction in clustering performance from 0.88 ($G=2$) to 0.73 ($G=20$), coupled with an increase in average runtime (2.47 to 197.31 seconds, respectively). Interestingly though, VSCC's performance remains remarkably consistent, and arguably improves (ignoring $G=2$) as the number of groups increases to 20 (due to a reduction in variation). However, VSCC does suffer a similar fate in runtime due to its reliance on {\tt mclust}.

Several things stand out in the results presented in Table~\ref{ta:g2}. For one, the linear relationship (which is also the most stringent of those considered) performs terribly under this simulation. Fortunately, the rest of the relationships put up solid performances and, in fact, very similar performances in general, as three of the four select six variables most often. Interestingly, no one relationship on its own would outperform {\tt mclust} on the full data set via either mean ARI or standard deviation; by choosing the best relationship via the total model uncertainty, however, the VSCC algorithm does narrowly beat out the full data set in both categories. This lends support to the use of uncertainty as a selection method. This is not the only simulation where we see results such as this, but it is not universally true across all simulations. We note further support for using the uncertainty while discussing the real data applications in our concluding paragraphs.

\subsection{Classification Example}
\begin{figure}[h]
\begin{center}
\begin{tabular}{cc}
\includegraphics[width=2.2in]{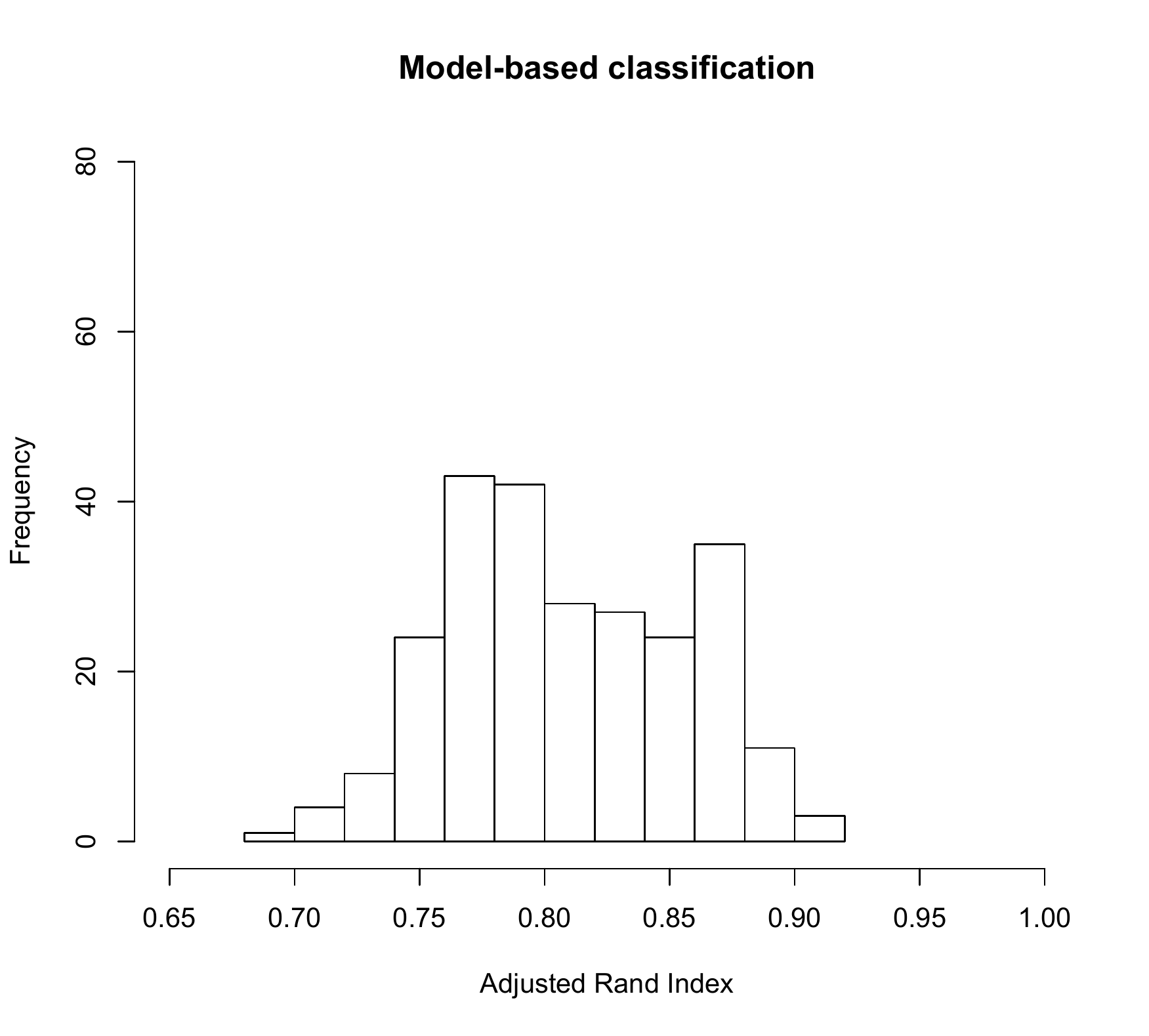} & \includegraphics[width=2.2in]{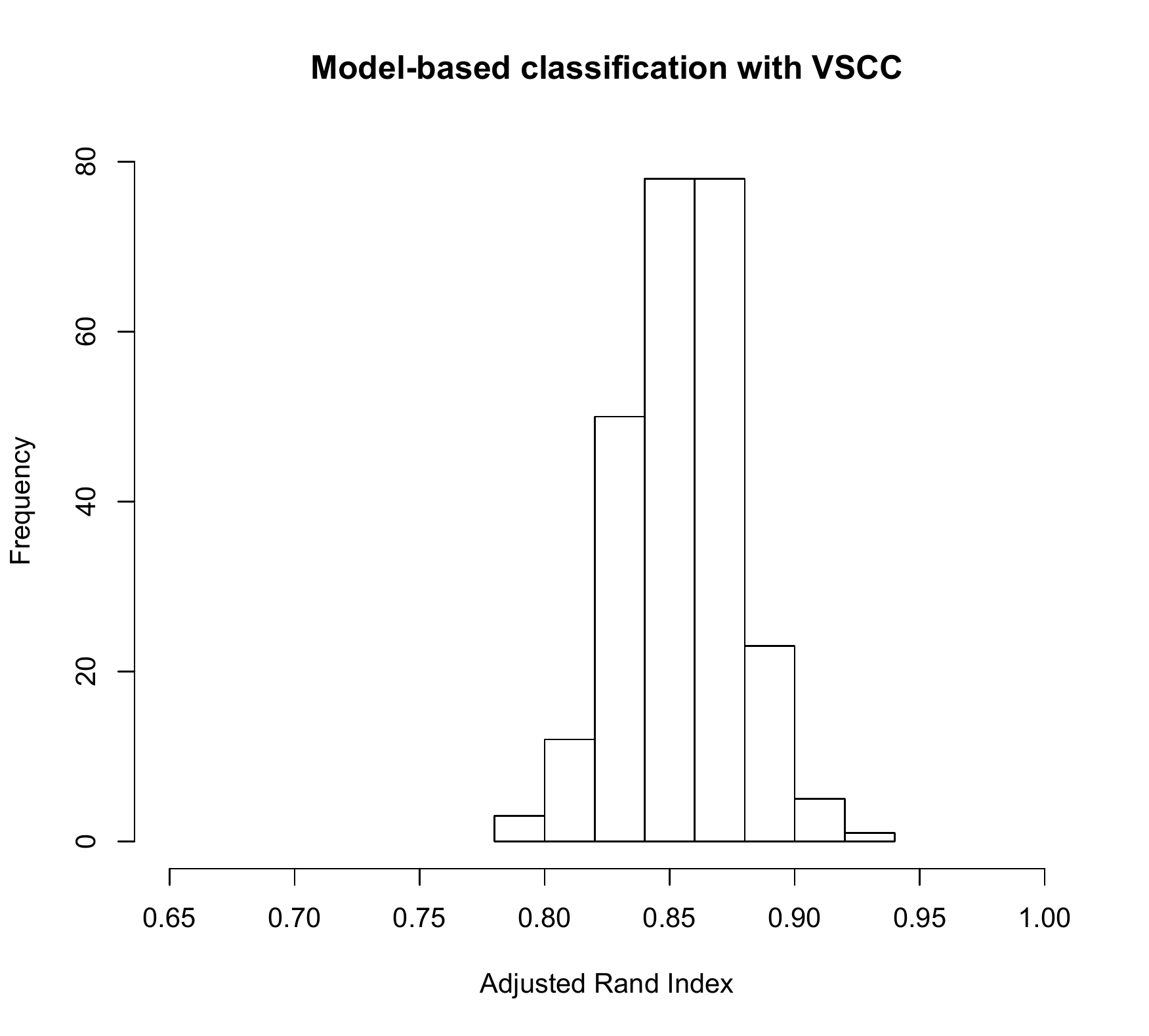}
\end{tabular}
\caption{Histograms of classification performance on the $G=15$ simulated data set by model-based classification and model-based classification with a reduced feature set selected by VSCC.}\label{fig:hists}
\end{center}
\end{figure}

To briefly demonstrate the feasibility of VSCC in a classification scenario, we apply the method under the supervised algorithm (cf. Section~\ref{sec:sup}) to the $G=15$ simulated data from the previous section. For each data set, we randomly select 50\% of the data to have known membership and analyze using model-based classification with the MCLUST family of models, then compare these results to using VSCC (with the same model-based classification on the chosen variables). This is performed on the 250 data sets, and a summary of classification performance can be deduced through Figure~\ref{fig:hists}. Note that the ARI reported only considers the observations with `unknown' cluster membership.

The mean ARI for analysis on the full data set is 0.81 with a 0.05 standard deviation, while the VSCC reduced data set achieves a mean of 0.85 with a 0.02 standard deviation. Recall the $G=15$ simulated data set contains 10 non-noisy and 10 noisy variables. VSCC (under the supervised algorithm with 50\% known) picks out the 10 meaningful variables 226 times, or for over 90\% of the data sets considered.

\section{Applications}\label{sec:app}
\subsection{Introduction}
The VSCC method will now be applied to  
real data sets under a clustering framework. An introduction to each of the four data sets is given at the beginning of each subsection. To facilitate interpretation, VSCC will be compared to the popular variable selection method introduced by \cite{raftery06}, available as the {\tt clustvarsel} package in {\sf R}, as well as the {\tt selvarclust} technique introduced by \cite{maugis09}, available as a command-line addition to the MIXMOD software \citep{biernacki06}. Recall from Section~\ref{sec:clus}, we utilize VSCC under a {\tt mclust} framework, meaning we use {\tt mclust} to initialize the $\hat z_{ig}$ and to cluster the feature-reduced data sets. In VSCC, we will utilize {\tt mclust} under the default settings: {\tt mclust} considers $G=1,\ldots,9$. Hence, for fairness of comparison, {\tt clustvarsel} and {\tt selvarclust} will be set to consider $G=1,\ldots,9$ as well. Note that all methods will be run on standardized variable (mean 0, variance 1) versions of the data discussed. Finally, {\tt selvarclust} will be restricted to the covariance parameterizations available in {\tt mclust}, again for fairness of comparison. Even so, the results presented from {\tt selvarclust} are not as directly comparable to the other two variable selection methods due to a difference in initializations used by the MIXMOD software versus {\tt mclust}.

\subsection{Leptograpsus Crabs}
The Leptograpsus crabs data set can be found in the {\tt MASS} library in {\sf R}. It contains five length measurements on two different colour forms of the crabs, further separated into the two genders. The results for the {\tt mclust} initialization, VSCC, {\tt clustvarsel}, and {\tt selvarclust} on the crabs data set are given in Table~\ref{ta:crabs}. 
\begin{table}[h]
\begin{center}
\caption{Table of results from {\tt mclust}, VSCC, {\tt clustvarsel}, and {\tt selvarclust} on the crabs data set. The relationship chosen for VSCC by the total model uncertainty is given in parentheses.}
\label{ta:crabs}
\begin{tabular}{lrrrrrr}
\hline
Analysis&ARI&Time (sec.)&$G$&\#Vars&Unc.\\
\hline
{\tt mclust}&0.31&3.94&4&5&14.71\\
VSCC (Quintic)&0.76&12.49&5&4&10.96\\
{\tt clustvarsel}&0.76&63.01&5&4&10.96\\
{\tt selvarclust}&0.50&256.69&5&4&12.79\\
\hline
\end{tabular}
\end{center}
\end{table}

On this data set, {\tt clustvarsel} (via approximate Bayes factors) and VSCC (via total uncertainty) agree on the solution that eliminates one variable and increases the ARI from 0.31 to 0.76. In fact, the {\tt selvarclust} algorithm selects the same variables, but the results are very different due to the initializations used. We can, for all intents and purposes, consider the performance of all techniques equivalent. The main item of interest here is that VSCC accomplishes this task five times faster than {\tt clustvarsel} and over 20 times faster than {\tt selvarclust}.

Perhaps a more important aspect of this application is that VSCC manages to increase clustering performance from a poor initialization (increasing the ARI from 0.31 for the correct number of groups to 0.76 for group over-estimation). One argument against an approach such as VSCC could be that one might need `quite good' initializations for the technique to be useful; the crabs data set, however, shows that this is not necessarily the case. Note that {\tt mclust} actually chooses the correct number of groups, meaning that the ARI is not artificially deflated by the choice of large $G$; its clustering performance is simply poor.

\subsection{Italian Wine}\label{sec:wine}
The Italian wine data set is readily available in the {\tt gclus} library in {\sf R} and contains 13 chemical measurements on 178 samples of wine originating from three different varieties (Barolo, Grignolino, and Barbera). From the results (Table~\ref{ta:wine}), we can see that VSCC outperforms the clustering done by {\tt mclust} alone as well as those done by {\tt clustvarsel} and {\tt selvarclust}. Running {\tt mclust} on the variables selected by {\tt selvarclust} results in the same performance as listed for {\tt selvarclust}. In addition to outperforming {\tt clustvarsel} and {\tt selvarclust} with respect to clustering performance, VSCC runs over 10 and 200 times faster, respectively.
\begin{table}[h]
\begin{center}
\caption{Table of results from {\tt mclust}, VSCC, {\tt clustvarsel}, and {\tt selvarclust} on the wine data set. The relationship chosen for VSCC by the uncertainty is given in parentheses.  }
\label{ta:wine}
\begin{tabular}{lcrcccc}
\hline
Analysis&ARI&Time (sec.)&$G$&\#Vars&Unc.\\
\hline
{\tt mclust}&0.48&1.91&8&13&5.85\\
VSCC (Quartic)&0.90&10.25&3&9&0.90\\
{\tt clustvarsel}&0.78&113.95&3&5&2.23\\
{\tt selvarclust}&0.54&2220.42&7&8&6.35\\
\hline
\end{tabular}
\end{center}
\end{table}

\subsection{Swiss bank notes data}
The Swiss bank notes data set is also available in the {\tt gclus} library in {\sf R} and contains six measurements on 200 monetary bills, of which some are legal tender and others counterfeit. The results (Table~\ref{ta:bank}) show that VSCC again results in the best clustering performance, with an ARI of 0.85 compared to 0.68 and 0.67 for {\tt mclust} and {\tt clustvarsel}, respectively. Running {\tt mclust} on the variables chosen by {\tt selvarclust} results in an ARI of 0.69, leaving it roughly on par with the {\tt mclust} and {\tt clustvarsel} results. VSCC utilizes fewer variables (4 versus 5) and runs eight times faster than {\tt clustvarsel}.
\begin{table}[h]
\begin{center}
\caption{Table of results from {\tt mclust}, VSCC, {\tt clustvarsel}, and {\tt selvarclust} on the bank notes data set. The relationship chosen for VSCC by the uncertainty is given in parentheses.}
\label{ta:bank}
\begin{tabular}{lcrccrr}
\hline
Analysis&ARI&Time (sec.)&$G$&\#Vars&Unc.\\
\hline
{\tt mclust}&0.68&2.34&4&6&6.16\\
VSCC (Quadratic)&0.85&8.52&3&4& 0.17\\
{\tt clustvarsel} &0.67&66.18&4&5&6.10\\
{\tt selvarclust} & 0.25 & 357.51 & 8 & 3 & 15.71 \\
\hline
\end{tabular}
\end{center}
\end{table}

\subsection{Coffee data}\label{sec:coff}
The coffee data set given by \cite{streuli73} contains 13 chemical measurements on 43 samples of coffee hailing from one of two species: Arabica or Robusto.These results (Table~\ref{ta:coff}) serve as an example where a variable selection method can negatively affect clustering performance. While {\tt mclust} performs perfect classification on the full data set, {\tt clustvarsel} and {\tt selvarclust} (including under {\tt mclust} analysis of the selected variables) select too many groups. 
\begin{table}[h]
\begin{center}
\caption{Table of results from {\tt mclust}, VSCC, {\tt clustvarsel}, and {\tt selvarclust} on the coffee data set. The relationship chosen for VSCC by the uncertainty is given in parentheses.}
\label{ta:coff}
\begin{tabular}{lcrcccc}
\hline
Analysis&ARI&Time (sec.)&$G$&\#Vars&Unc.\\
\hline
{\tt mclust}&1.00&0.16&2&13&0.00\\
VSCC (Quadratic)&1.00&1.19&2&2&0.00\\
{\tt clustvarsel}&0.41&2.79&3&6&0.42\\
{\tt selvarclust}&0.37&404.67&4&7&0.23 \\
\hline
\end{tabular}
\end{center}
\end{table}

\begin{figure}[h]
\begin{center}
\includegraphics[width=2.5in]{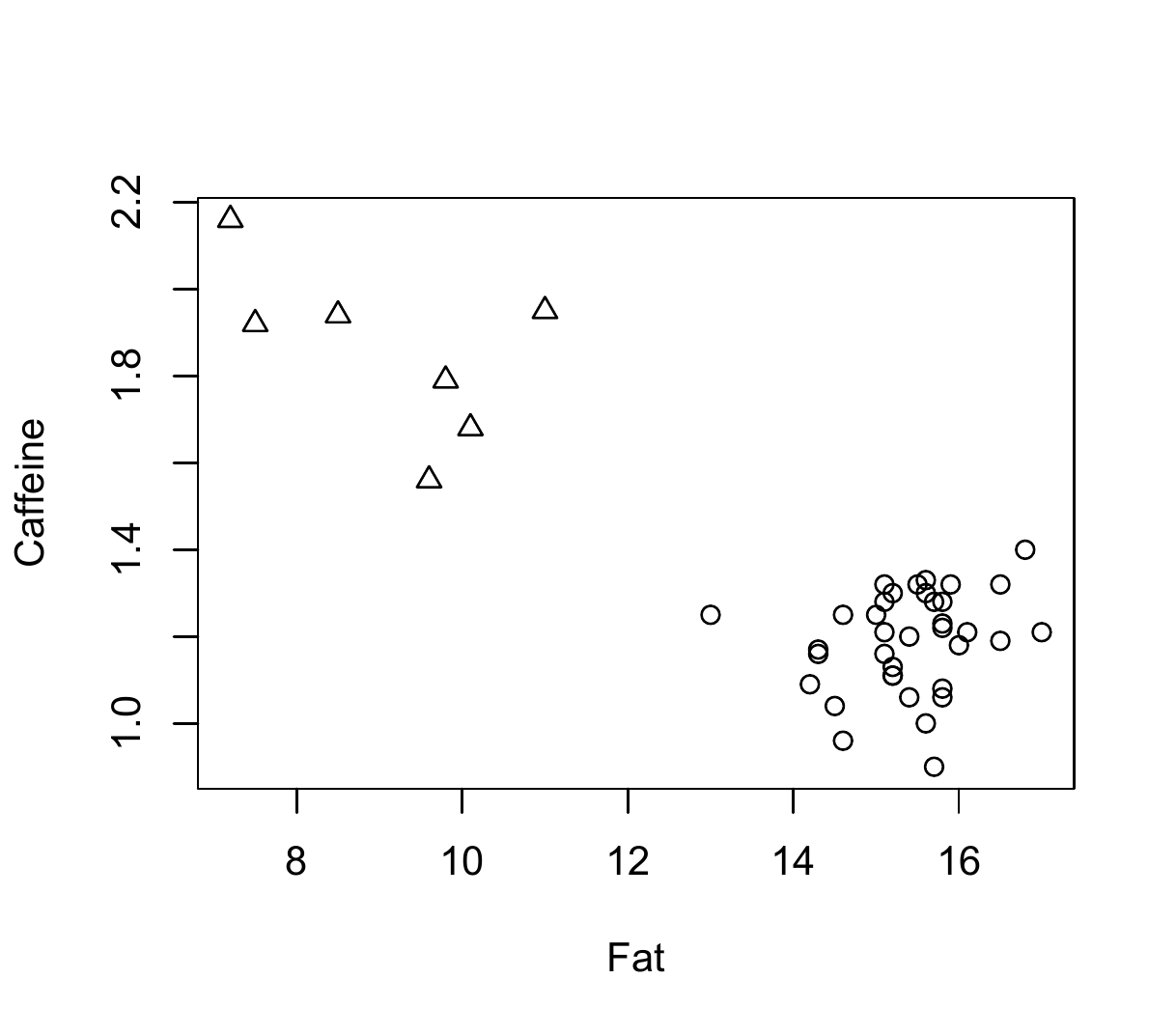}
\caption{Scatterplot of the Arabica (circles) and Robusto (triangles) coffee beans on the VSCC chosen variables.}\label{fig:coff}
\end{center}
\end{figure}

Perhaps more importantly, VSCC gives perfect classification while reducing the number of variables from 13 to 2: caffeine and fat content. A visualization of the clusters on these two variables is given in Figure~\ref{fig:coff}. 

\subsection{Comparison with Correctly Specified $G$}\label{sec:skm}
We conclude the analysis of real data sets by comparing VSCC with results from the {\tt sparcl} package \citep{sparcl} in {\sf R}, which contains an implementation of the sparse $k$-means technique described by \cite{witten10}. We use this approach under default settings, allowing {\tt KMeansSparseCluster.permute()} to select the tuning parameter. Because this approach is based on $k$-means, the number of groups must be specified; this thus differs from the approaches compared in the previous sections. Table~\ref{ta:skm} contains the results from all four data sets analyzed by {\tt sparcl} and VSCC.
\begin{table}[h]
\begin{center}
\caption{Table of results from VSCC and {\tt sparcl} on the previously introduced data sets with the correct number of groups pre-specified. The relationship chosen for VSCC by the uncertainty is given in parentheses.}
\label{ta:skm}
\begin{tabular}{ccccccc}
\hline
Data Set & Analysis&ARI&Time (sec.)&\#Vars\\
\hline
\multirow{2}{*}{Crabs} & VSCC (Quintic)&0.37&1.01&4\\
& {\tt sparcl}&0.02&6.17&5\\
\hline
\multirow{2}{*}{Wine} & VSCC (Cubic)&0.93&0.33&7\\
& {\tt sparcl}&0.85&6.36&13\\
\hline
\multirow{2}{*}{Bank} & VSCC (Quintic)&0.98&0.26&5\\
& {\tt sparcl}&0.96&4.55&6\\
\hline
\multirow{2}{*}{Coffee} & VSCC (Cubic)&1.00&0.12&6\\
& {\tt sparcl}&1.00&2.71&13\\
\hline
\end{tabular}
\end{center}
\end{table}

The results across all four data sets show that VSCC can be run in a significantly shorter amount of time and result in better (or equal, in one case) clustering performance. Note also that {\tt sparcl} has reduced the dimensionality of the data set but has not technically removed any variables from consideration (because none of the variables have been weighted to zero in the model). As a small side note, we point out that the VSCC results on the coffee and wine data sets differ slightly from those earlier in Sections~\ref{sec:wine} and \ref{sec:coff}, even though the final number of groups selected is the same. This is due to differences in initializations from the preliminary {\tt mclust} runs, and thus is not particularly surprising.

\section{Discussion and Future Work}\label{sec:disc}
A novel variable selection technique (VSCC) based on within-group variance was introduced and utilized under a model-based clustering framework. The strengths of the technique lie in the speed at which it can be run as well as its intuitive appeal. It was shown to outperform or equal {\tt clustvarsel}, {\tt selvarclust}, and {\tt sparcl}  in clustering performance, and significantly outperform them in speed, on four commonly used data sets. Notably, the VSCC relationship chosen by the total uncertainty was, in every real data set considered, the relationship that resulted in the highest ARI. This, along with several of the simulation studies, lends support for the use of total uncertainty as a subset selection criteria.

The inner workings of VSCC are flexible enough to incorporate clustering/classification algorithms other than the model-based techniques covered in this article. One hurdle to overcome in this respect is an effective subset selection criterion --- as non-model-based methods will not contain uncertainty measures --- and this will be a subject of future research. In addition, the development of VSCC software for the {\sf R} computing environment is intended pending code optimization and further testing.  

\section*{Acknowledgements}
The authors wish to acknowledge helpful comments from the editor and two anonymous peer reviewers. This work was supported by a Postgraduate Doctoral Scholarship (Andrews) and a Discovery Grant (McNicholas) from the Natural Sciences \& Engineering Research Council of Canada; by an Early Researcher Award from the Ontario Ministry of Research \& Innovation (McNicholas); and by the University Research Chair in Computational Statistics at the University of Guelph (McNicholas).

\bibliographystyle{chicago}
\bibliography{bib}

\end{document}